\documentclass[journal]{IEEEtran}
\ifCLASSINFOpdf
   \usepackage[pdftex]{graphicx}
\else
\fi
\usepackage{multirow}
\usepackage{subfig}
\begin{document}
%
\title{A Hybrid Frequency-domain/Image-domain Deep Network for Magnetic Resonance Image Reconstruction}
%
%
%

\author{Roberto Souza,~\IEEEmembership{Member,~IEEE,}
        and~Richard Frayne 
\thanks{R Souza and R. Frayne are with the Departments of Radiology and Clincial Neuroscience, Hotchkiss Brain Institute, University of Calgary, Calgary, AB, T2N 1N4 Canada, and the Seaman Family MR Research Centre, Foothills Medical Centre, Alberta Health Services, Calgary, Alberta, T2N 2T9 Canada.e-mail: roberto.medeirosdeso@ucalgary.ca, rfrayne@ucalgary.ca}
\thanks{This work has been submitted to the IEEE for possible publication. Copyright may be transferred without notice, after which this version may no longer be accessible.}
\thanks{Manuscript received October 30, 2018;
}}

%
%

\markboth{arXiv preprint}
{Shell \MakeLowercase{\textit{et al.}}: Bare Demo of IEEEtran.cls for IEEE Journals}
%



\maketitle

\begin{abstract}
Decreasing magnetic resonance (MR) image acquisition times can potentially reduce procedural cost and make MR examinations more accessible. Compressed sensing (CS)-based image reconstruction methods, for example, decrease MR acquisition time by reconstructing high-quality images from data that were originally sampled at rates inferior to the Nyquist-Shannon sampling theorem. Iterative algorithms with data regularization are the standard approach to solving ill-posed, CS inverse problems. These solutions are usually slow, therefore, preventing near-real time image reconstruction. Recently, deep-learning methods have been used to solve the CS MR reconstruction problem. These proposed methods have the advantage of being able to quickly reconstruct images in a single pass using an appropriately trained network. Some recent studies have demonstrated that the quality of their reconstruction equals and sometimes even surpasses the quality of the conventional iterative approaches. A variety of different network architectures (\textit{e.g.}, U-nets and Residual U-nets) have been proposed to tackle the CS reconstruction problem. A drawback of these architectures is that they typically only work on image domain data. For undersampled data, the images computed by applying the inverse Fast Fourier Transform (iFFT) are aliased. In this work we propose a hybrid architecture that works both in the k-space (or frequency-domain) and the image (or spatial) domains. Our network is composed of a complex-valued residual U-net in the k-space domain, an iFFT operation, and a real-valued U-net in the image domain. Our experiments demonstrated, using MR raw k-space data, that the proposed hybrid approach can potentially improve CS reconstruction compared to deep-learning networks that operate only in the image domain. In this study we compare our method with four previously published deep neural networks and examine their ability to reconstruct images that are subsequently used to generate regional volume estimates. We evaluated undersampling ratios of 75\% and 80\%. Our technique was ranked second in the quantitative analysis, but qualitative analysis indicated that our reconstruction performed the best in hard to reconstruct regions, such as the cerebellum. All images reconstructed with our method were successfully post-processed, and showed good volumetry agreement compared with the fully sampled reconstruction measures. 
\end{abstract}

\begin{IEEEkeywords}
Compressed sensing (CS), magnetic resonance (MR), machine learning, convolutional neural network (CNN), image reconstruction
\end{IEEEkeywords}

%
\IEEEpeerreviewmaketitle
\section{Introduction}
\IEEEPARstart{M}{agnetic resonance} (MR) is a key medical imaging modality that has critical roles in both patient care and medical research. MR scanner installations, however, are expensive. Another major limitation to MR imaging is the comparatively long image acquisition times, especially when compared to other modalities like computerized tomography. Lengthy acquisition times make MR less patient friendly and increase the per patient examination cost. MR-based compressed sensing (CS) methods seek to leverage the implicit sparsity of medical images \cite{RN280}, potentially allowing for significant k-space undersampling during acquisition, and by consequence, reducing examination times. 

Traditional MR CS reconstruction techniques are iterative algorithms that usually require a sparsifying transform that when combined with regularization parameters are able to find a solution for these ill-posed inverse problems \cite{RN280,RN247}. The drawback of these iterative approaches, however, is that they are time-consuming, making them more difficult to incorporate in a near real-time MR imaging scenario (\textit{i.e.}, where images are reconstructed and displayed on the scanner during the procedure).

Deep learning \cite{RN250} is a new method that has been applied to reconstruction challenges. It has the advantage of being able to rapidly reconstruct images in a single-pass using a suitably trained network. Some deep-learning based reconstruction methods have arguably surpassed traditional iterative CS reconstruction techniques \cite{RN307}. 

A few different deep learning approaches have been recently proposed to tackle the CS reconstruction problem. Jin \textit{et al.} \cite{RN254} proposed to use a U-net \cite{RN196}. Lee  \textit{et al.} experimentally showed that residual U-nets can potentially improve image reconstruction. Residual blocks have subsequently been incorporated in the majority of the latest studies (\textit{cf.}, \cite{RN307,RN305,RN306}).

Yang \textit{et al.} \cite{RN307} proposed a deep de-aliasing generative adversarial network (DAGAN) that uses a residual architecture as the network generator responsible for reconstructing the images associated to a loss function that has four components: an image domain loss, a frequency domain loss, a perceptual loss, and an adversarial loss. Quan \textit{et al.} \cite{RN305} proposed a generative adversarial network (GAN) with a cyclic loss \cite{RN315}. Their method consists of a cascade of a reconstruction network followed by a refinement network with a cyclic loss component that tries to enforce that the model is bijective. Schlemper \textit{et al.} \cite{RN306} proposed a deep cascade of convolutional neural networks (CNNs) that has data consistency (DC) blocks between consecutive subnetworks in the cascade. Their hypothesis is that the DC blocks reduce the issue of overfitting, therefore allowing the training of deeper models. 

The aforementioned techniques principally work in the image domain, with a few exceptions, where k-space domain information is used in the loss function and/or to implement DC layers. All of these networks \cite{RN307,RN305,RN306} receive as input the undersampled k-space zero-filled reconstruction, and output an unaliased image. 

Zhu \textit{et al.} \cite{RN289} recently proposed a method that tries to learn the domain transform. Their method first processes the undersampled input data in k-space, learns the inverse discrete Fourier transform (iDFT), and then refines the result in the image domain. In the case of CS reconstruction, the domain transform can be considered as learning an approximation for the iDFT for undersampled k-space data. The domain transform is modeled as a sequence of connected layers, and the image domain refinement is modeled as a series of convolutional layers. A disadvantage of this approach is that the domain transform has a quadratic complexity with respect to the size of the input image. For example, when dealing with $256\times 256$ images, the number of learned parameters in their model would be greater $>10^{10}$. The quadratic order of the algorithm makes it challenging to use their model for typical MR image sizes due to current hardware limitations.

Based on these studies, we hypothesize that a hybrid approach that works with both information as presented in k-space domain and image domain (such as the proposal of \cite{RN289}) can improve MR CS reconstruction. In this work, we propose a model that consists of a cascade of a k-space domain network  and an image domain network connected by the magnitude of the iDFT operation. Our model does not need to learn the domain transform, which essentially reduces our model parameter complexity to $O(N^2)$. For a $256\times 256$ input image, the number of learned parameters is reduced by a factor of $10^3$ for our method compared to \cite{RN289} ($\approx 10^{7}$ parameters). Our proposed method takes advantage of information as presented in k-space and image domain, as opposed to other image domain only approaches \cite{RN250,RN307,RN254,RN305,RN306}. K-Space and image domain information are equivalent, because they are related by a global linear transformation. Operating in both domains with non-linear methods, however, can potentially be advantageous and improve the network learning. 

In this work, we compare our proposal against four recently published, publicly available, deep learning-based reconstruction methods \cite{RN307,RN254,RN305,RN306}. These approaches were evaluated at $75\%$ and $80\%$ undersampling levels (corresponding to acquisition acceleration factors of $4\times$ and $5\times$). The experiments were done using MR raw-data acquired from subjects scanned using a volumetric $T1$-weighted MR imaging sequence. The proposed reconstruction code has been made publicly available  at  \underline{https://github.com/rmsouza01/Hybrid-CS-Model-MRI}. The dataset will be made publicly available for benchmarking purposes at \underline{https://sites.google.com/view/calgary-campinas-dataset/home} and is part of the Calgary-Campinas datasets \cite{RN136}. 

\section{Hybrid Network}
\begin{figure*}[!ht]
\centering
\includegraphics[width=1.0\textwidth]{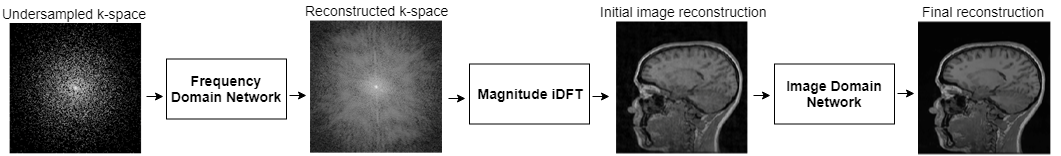}
\caption{Flowchart of the proposed methodology. The frequency domain network can be seen as k-space interpolation to fill the missing values. The image domain network acts as an anti-aliasing filter to further improve the image reconstruction obtained from the first network.}
\label{fig_flowchart}
\end{figure*}

The flowchart of our proposed method is depicted in Figure \ref{fig_flowchart}. There are three main components to our approach: 1) a k-space (or frequency-domain) network that essentially tries to fill missing k-space samples 2) the magnitude of the iDFT, and 3) an image-domain network that acts as an anti-aliasing filter. These components along with the proposed network loss function are described in the following subsections.

\subsection{Frequency-domain Network}
The frequency-domain network, $f_{cnn_1}$, attempts to recover a fully sampled k-space, $\widehat{F}_{norm}(k_x,k_y)$, given undersampled k-space data, $F_{u}$. This can be mathematically formalized as:

\begin{equation}
\widehat{F}_{norm}(k_x,k_y) = f_{cnn_1}[F_{u_{norm}}],
\end{equation}

\noindent where $F_{u_{norm}}$ is the normalized undersampled k-space data given by:

\begin{equation}
F_{u_{norm}} = \frac{F_{u} - \mu_{F_{u_{train}}}}{\sigma_{F_{u_{train}}}},
\end{equation}

\noindent where $\mu_{F_{u_{train}}}$ and $\sigma_{F_{u_{train}}}$ are the average and standard-deviation of undersampled k-spaces in the training set.
The specific architecture used for the frequency-domain network is a residual U-net (Figure \ref{fig_architecture}, left side). The input complex k-space image is split in two-channels: one for the real and other for the imaginary components of the k-space data.

\subsection{Magnitude of the iDFT}
Before applying the iDFT, we have to undo the previous k-space normalization step. Adding a constant to the k-space data results in superposition of an impulse, $\delta(\cdot)$, signal to the image after transformation.
Undoing the normalization is accomplished by:
 
\begin{equation}
\widehat{F}(k_x,k_y) = F_{u_{norm}}\times\sigma_{F_{u_{train}}}  + \mu_{F_{u_{train}}}.
\end{equation}

\noindent The next step is to transform from the frequency domain to the image domain using the iDFT and magnitude operations:

\begin{equation}
\widehat{f_0} = ||\mathcal{F}^{-1}(\widehat{F})||
\end{equation}

\noindent where $\mathcal{F}^{-1}$ represents the iDFT operation and $\widehat{f_0}$ is the initial estimate of the reconstructed image. 

This component of our model has no trainable parameters, and the iDFT is efficiently computed using the fast Fourier transform algorithm running on a graphics processing unit (GPU).

\subsection{Image Domain Network}
The last component of our method is the image domain network ($f_{cnn_2}$). In order to improve training convergence of the network, we again normalize the initial estimate of the reconstructed image obtained in the previous step:

\begin{equation}
\widehat{f}_{0_{norm}} = \frac{\widehat{f_0} - \mu_{f_{0_{train}}}}{\sigma_{f_{0_{train}}}}.
\end{equation}
\noindent where $\mu_{f_{0_{train}}}$ and $\sigma_{f_{0_{train}}}$ are the mean and standard-deviation of the reconstructed images in the training set.
The normalized image $\widehat{f}_{0_{norm}}$ is fed as input to the image domain network:

\begin{equation}
\widehat{f}(x,y) = f_{cnn_2}[\widehat{f}_{0_{norm}}].
\end{equation}

\begin{figure*}[!ht]
\centering
\includegraphics[width=1.0\textwidth]{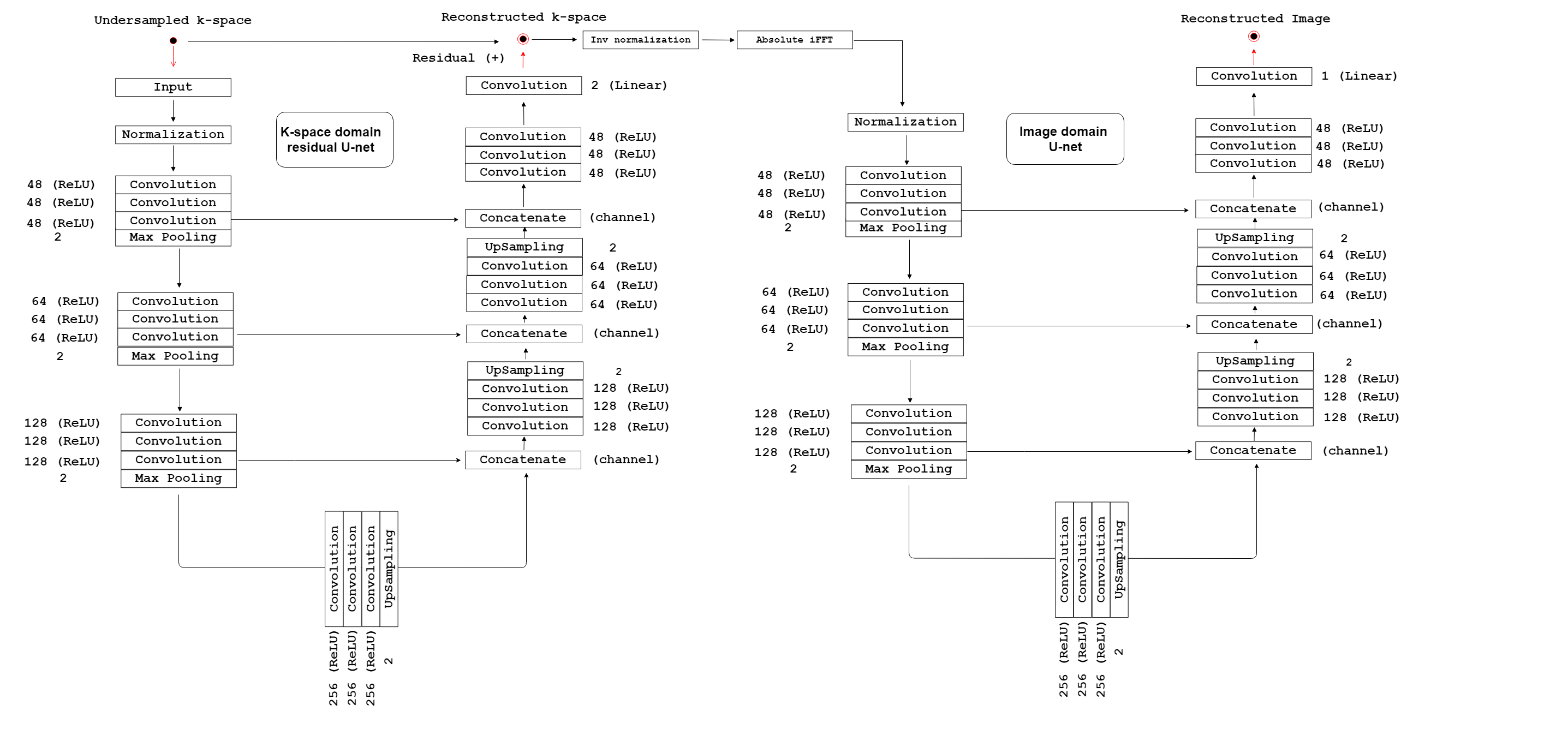}
\caption{Architecture of our Hybrid model. The k-space network uses a $5 \times 5$ convolution kernel size and the image domain network uses a $3 \times 3$ kernel.}
\label{fig_architecture}
\end{figure*}

\noindent The final reconstructed image is $\widehat{f}(x,y)$. The architecture used for the image domain network is a U-net (Figure \ref{fig_architecture}, right side). 

\subsection{Loss Function}
Our loss function was a weighted sum of normalized root mean squared errors (NRMSE) in each domain given by:

\begin{equation}
\mathcal{L} = \frac{1}{N}\sum^N_{i=1} w_1\times NRMSE(F_i,\widehat{F_i}) + w_2\times NRMSE(f_i,\widehat{f_i}),
\end{equation}

\noindent where $F_i$ and $f_i$ are the reference fully sampled k-space and image reconstruction, respectively, of the $i$-th sample in the training set, and $N$ is the number of training samples. The first term of the loss function acts as a data fidelity term, \textit{i.e.}, a regularizer, and the second term represents the reconstruction error. In our experiments $w_1=0.001$ and $w_2=0.999$. These values were empirically determined. 

\section{Experimental Setup}
\subsection{Network Implementations}
We compared our method, which we will refer to as the Hybrid method, against 1) a plain vanilla U-net \cite{RN196} with a residual connection, referred to as UNET; 2) RefineGAN \cite{RN305}; 3) DAGAN \cite{RN307}; and 4) Deep-Cascade \cite{RN306} with a cascade of five CNNs and five DC blocks, which is the network configuration that the authors reported best results.

We used the Keras application program interface \cite{chollet2015keras} using TensorFlow as backend \cite{tensorflow2015-whitepaper} to implement our hybrid network and the UNET. For RefineGAN, DAGAN and Deep-Cascade, we used the source code provided by the authors. All networks were trained using our data for acceleration factors of $4\times$ and $5\times$ using two-dimensional Gaussian undersampling patterns. The networks were trained and tested on Amazon Elastic Compute Cloud services using a p3.2xlarge\footnote{https://aws.amazon.com/ec2/instance-types/p3/} instance, which has a NVIDIA Tesla V100 GPU.

\subsection{Training, Validation and Testing Dataset}
Our dataset consists of 45 volumetric $T1$-weighted, fully sampled k-space datasets acquired on a clinical MR scanner (Discovery MR750; General Electric (GE) Healthcare, Waukesha, WI) that were collected as part of the ongoing Calgary Normative Study \cite{tsang2017white}.
The data was acquired with a $12$-channel imaging coil and an acquisition matrix of size $252\times 256$.
Data were zero-filled to an image matrix size of $256 \times 256$. The multi-coil k-space data was reconstructed using vendor supplied tools (Orchestra Toolbox; GE Healthcare). Coil sensitivity maps were normalized to produce a single complex-valued image set that could be back-transformed to regenerate complex k-space samples. In our experiments, we performed retrospective undersampling, effectively simulating a single-coil acquisition scenario. Our train/validation/test data split was 25/10/10,  equivalent to 4,524 slices/1,700 slices/1,700 slices. There was no overlap of same subject slices in the train, validations and test sets.

\subsection{Performance Metrics}
The performance metrics used to assess the networks were:
\begin{itemize}
\item NRMSE
\begin{equation}
NRMSE(\widehat{f},f) = \frac{\sqrt{\frac{1}{M}\sum^M_{i=1} [\widehat{f}(i)-f(i)]^ 2}}{max(f) - min(f)},
\label{nrmse}
\end{equation}
where $M$ is the number of pixels (or voxels) in the image.
\item Structural Similarity (SSIM) \cite{RN316}
\begin{equation}
SSIM(\widehat{f},f) = \frac{(2\mu_f\mu_{\widehat{f}}+c_1)(2\sigma_{f\widehat{f}}+c_2)}{(\mu_f^2 + \mu_{\widehat{f}}^2+c_1)(\sigma_f^2 + \sigma_{\widehat{f}}^2+c_2)},
\end{equation}
where $c_1$ and $c_2$ are two variables used to stabilize the division and $\mu$ and $\sigma^2$ represent mean and variance values of the gray-level intensities of the images.
\item Peak Signal to Noise Ratio (PSNR):
\begin{equation}
PSNR(\widehat{f},f) = 20log_{10}(\frac{max(f)}{\sqrt{\frac{1}{M}\sum^M_{i=1}[\widehat{f}(i)-f(i)]^ 2}}).
\end{equation}
\end{itemize}

These metrics were used because they are commonly used to assess image reconstruction. The higher the SSIM and PSNR values the better the result. For NRMSE, smaller values represent better reconstructions.

\subsection{Volumetric Assessment}
For the top two performing reconstruction techniques, we performed volumetric analysis  with a commonly used software for neuroimaging analysis (FreeSurfer \cite{fischl2012freesurfer}). We considered the fully sampled reconstruction results as our gold-standard. Only the ten test volumes were analyzed.
We recorded number of processing failures and analyzed the average absolute deviation of total intra-cranial, white-matter, gray-matter, hippocampus, and amygdala volumes.

\subsection{Statistical Methods}
We report mean and standard deviation of all average measures. We used a one-way analysis of variance (ANOVA) to determine statistically significant changes and \textit{post-hoc} paired \textit{t}-tests to determine statistically significant pair-wise differences. A $p$-value $<0.01$ was used to determine statistical significance.

\section{Results}
\subsection{Quantitative Assessment}
The reconstruction metrics are summarized in Table \ref{table_results}. Deep-Cascade achieved the best quantitative results followed by our Hybrid approach across all performance metrics. The one-way ANOVA tests showed statistically significant differences ($p<0.01$) across all metrics and acceleration factors.
The paired \textit{t}-tests showed that Deep-Cascade was statistically better ($p<0.01$) than the other methods in the comparison and our Hybrid, while worse than Deep-Cascade, was statistically better than UNET, DAGAN, and RefineGAN ($p<0.01$). We note that the absolute differences between Deep-Cascade and Hybrid networks, while statistically significant, were small.

\begin{figure*}
\centering 
\subfloat[Fully sampled]{\includegraphics[width=0.275\textwidth]{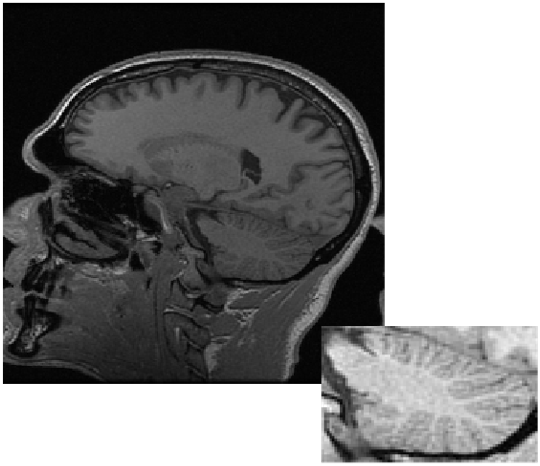}}
\subfloat[UNET]{\includegraphics[width=0.275\textwidth]{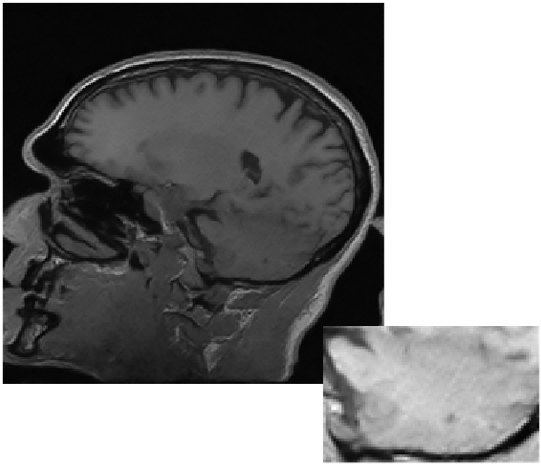}}
\subfloat[DAGAN]{\includegraphics[width=0.275\textwidth]{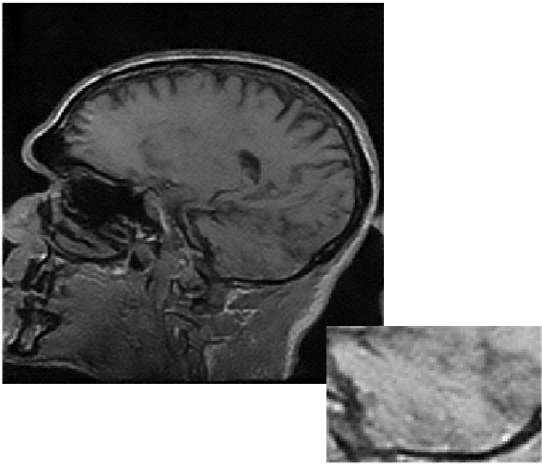}}\\
\subfloat[RefineGAN]{\includegraphics[width=0.275\textwidth]{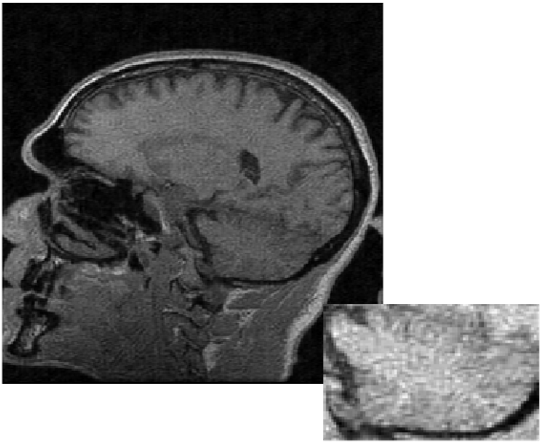}}
\subfloat[Deep-Cascade]{\includegraphics[width=0.275\textwidth]{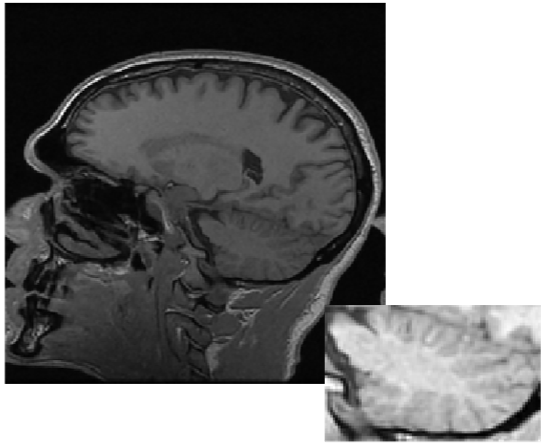}}
\subfloat[Hybrid]{\includegraphics[width=0.275\textwidth]{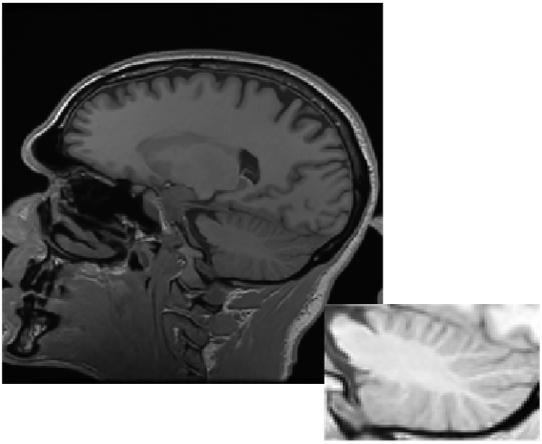}}
\caption{Sample reconstructions with a special highlight on the cerebellum region, where differences are more noticeable.}
\label{reconstruction}
\end{figure*}

The average NRMSE across the slices in the ten test volumes is depicted in Figure \ref{nrmse}. Over the central slices NMRSE followed a predictable pattern. Towards the edges, where little of the brain was present, the error increased. Sample reconstructions for each technique are shown in Figure \ref{reconstruction}. Qualitatively the Hybrid network produces the most pleasing image, particularly in regions with large signal differences like the cerebellum.

\begin{figure}
\centering
\includegraphics[width=0.4\textwidth]{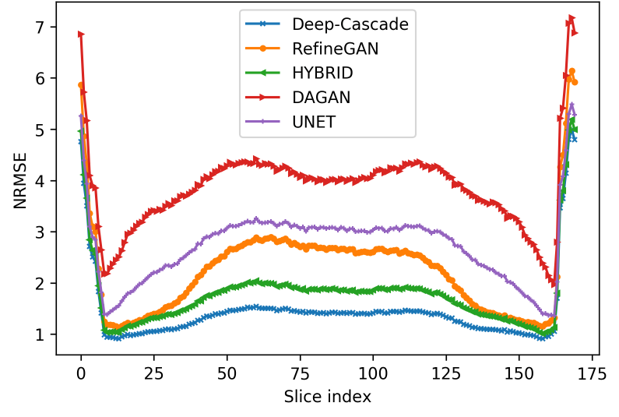}
\caption{Average NRMSE distribution along the slices. Edge slices that do not have much signal have higher average errors.}
\label{nrmse}
\end{figure}

\subsection{Volumetric Analysis}
\begin{figure*}
\centering
\includegraphics[width=0.9\textwidth]{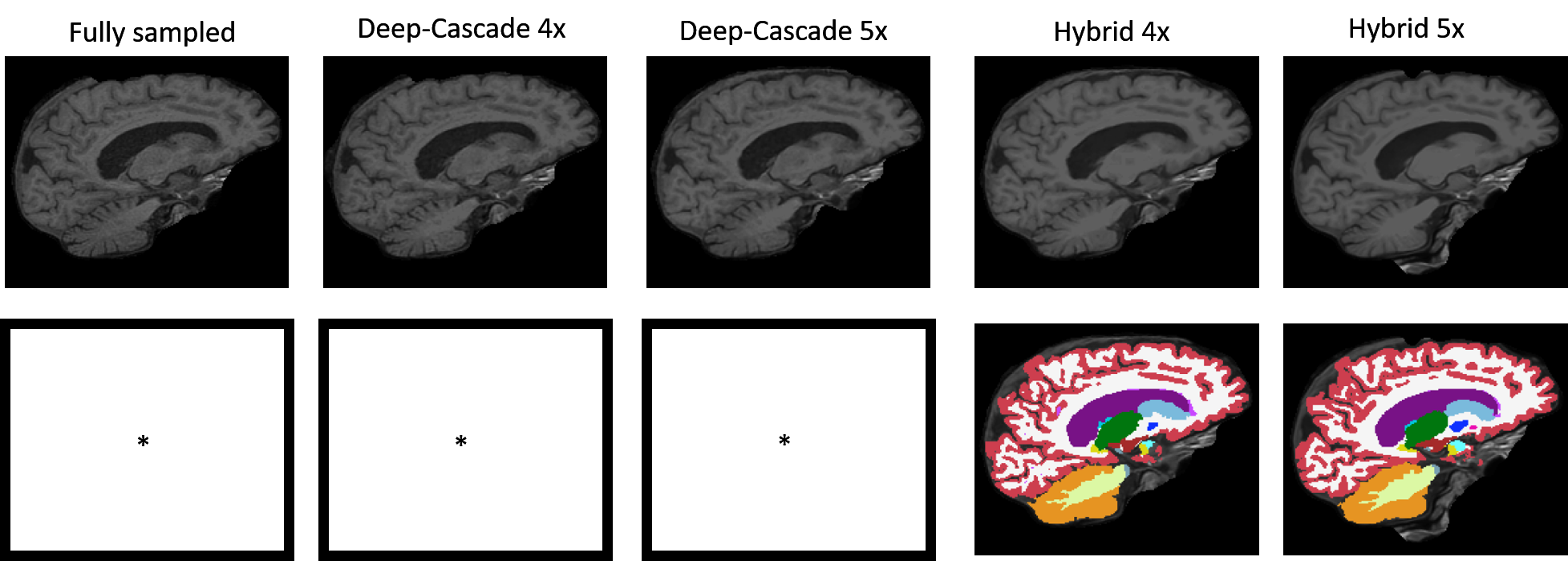}
\caption{Sample result of volumetric analysis failures in one subject. Processing failed for the fully sampled reconstruction, and Deep-Cascade (acceleration factors of $4\times$ and $5\times$). * represents an analysis that failed.}
\label{freesurfer_failed}
\end{figure*}

\begin{figure*}[!ht]
\subfloat[]{\includegraphics[width=0.25\textwidth]{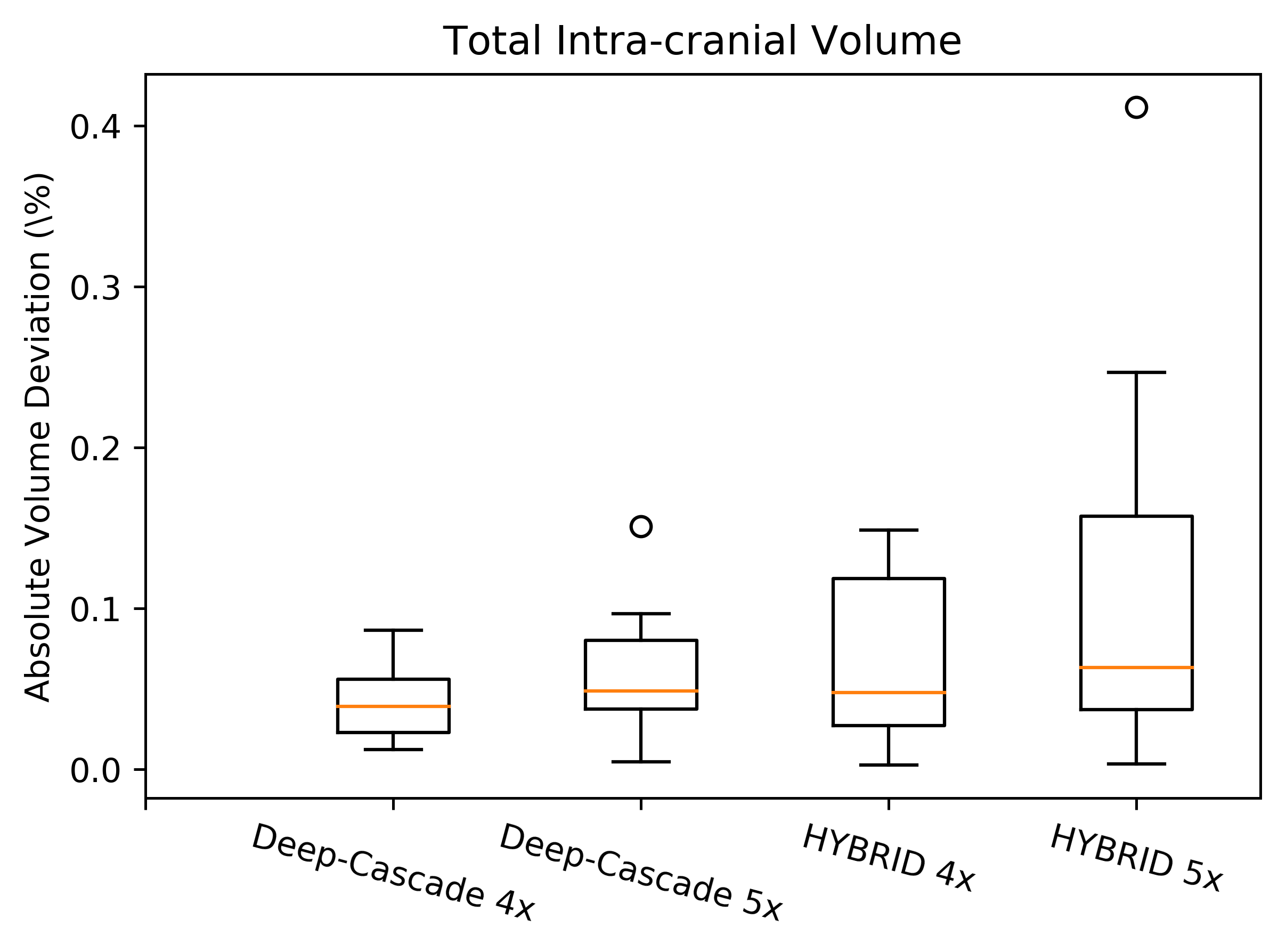}}
\subfloat[]{\includegraphics[width=0.25\textwidth]{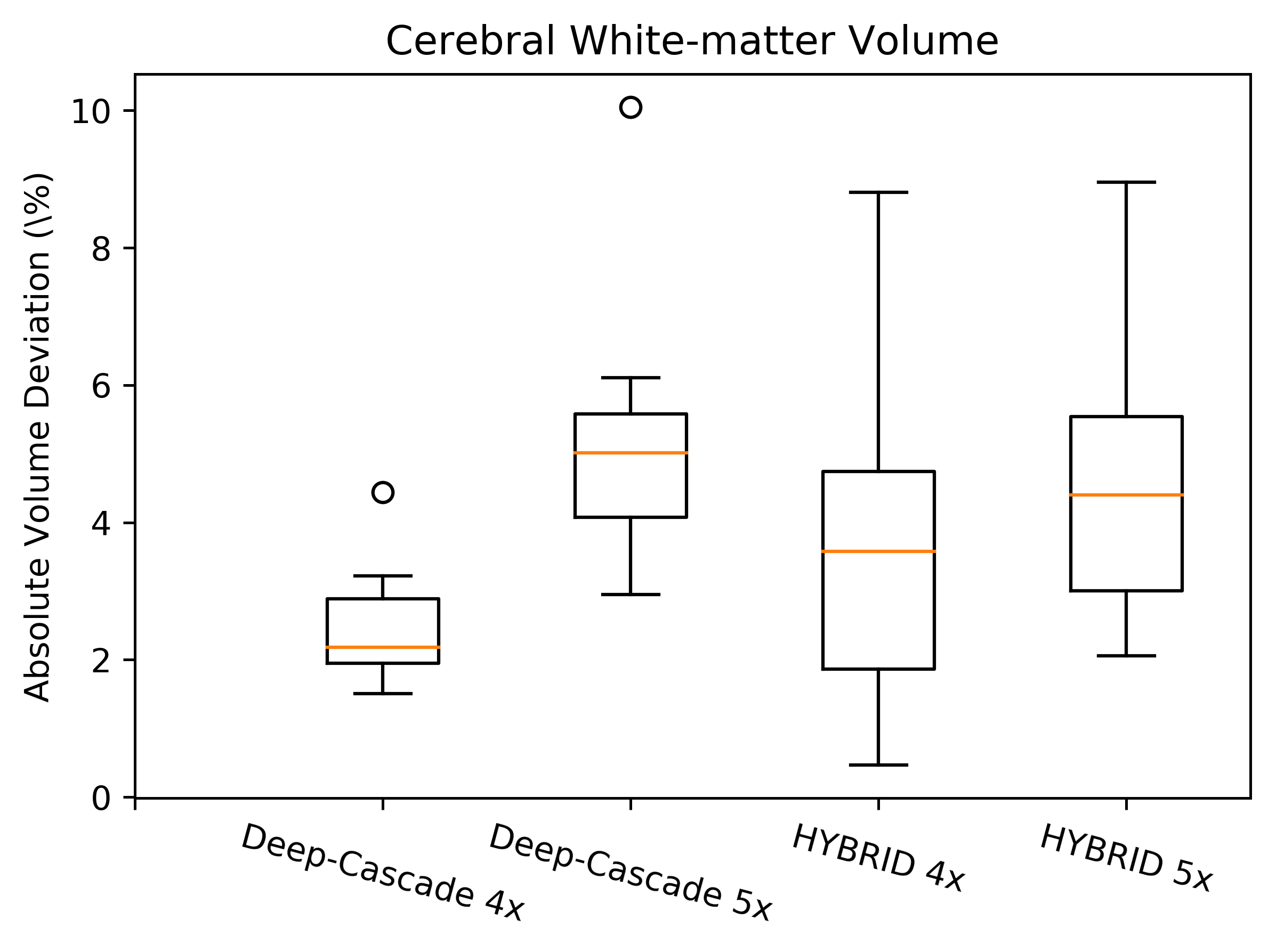}}
\subfloat[]{\includegraphics[width=0.25\textwidth]{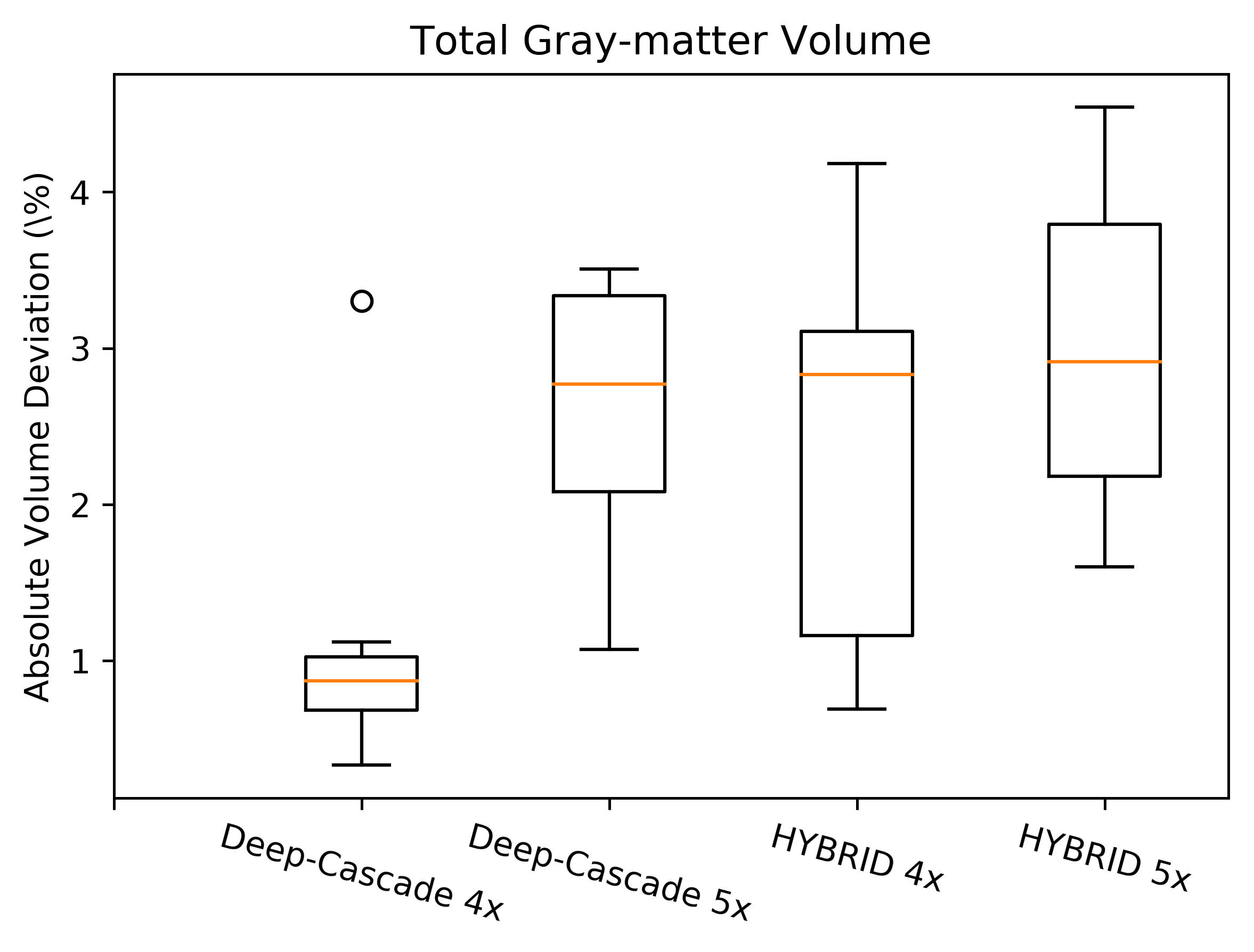}}
\subfloat[]{\includegraphics[width=0.25\textwidth]{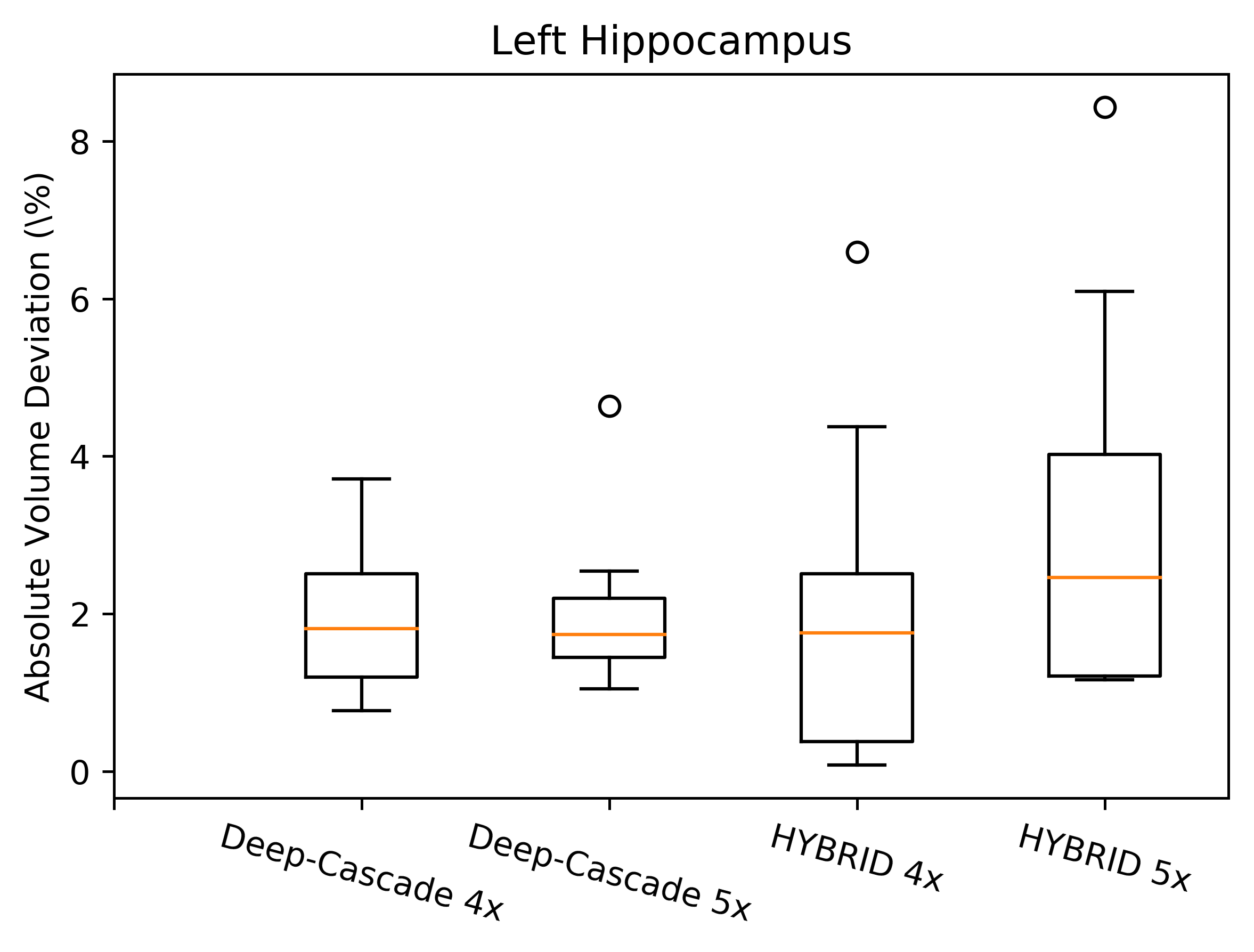}}\\
\centering 
\subfloat[]{\includegraphics[width=0.25\textwidth]{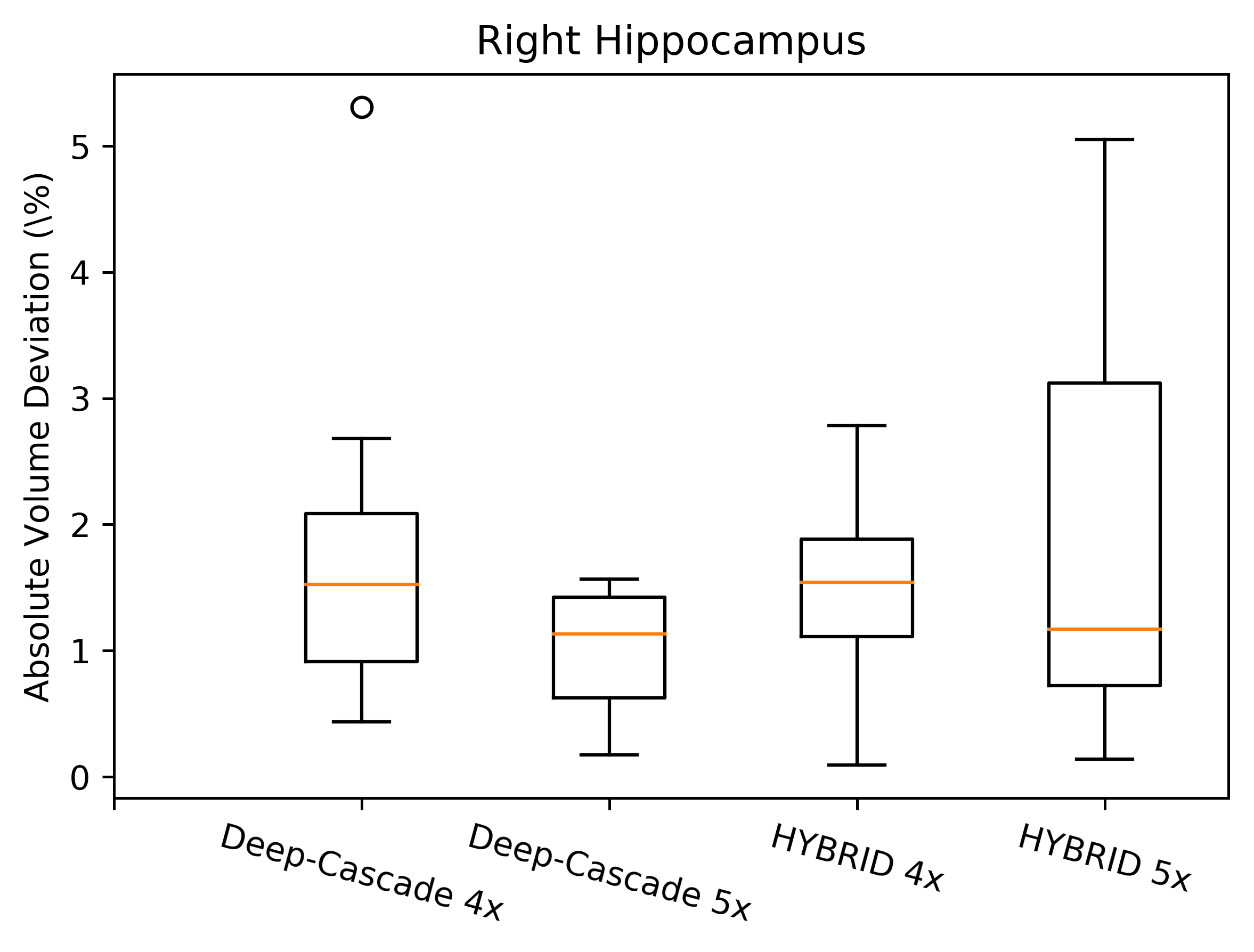}}
\subfloat[]{\includegraphics[width=0.25\textwidth]{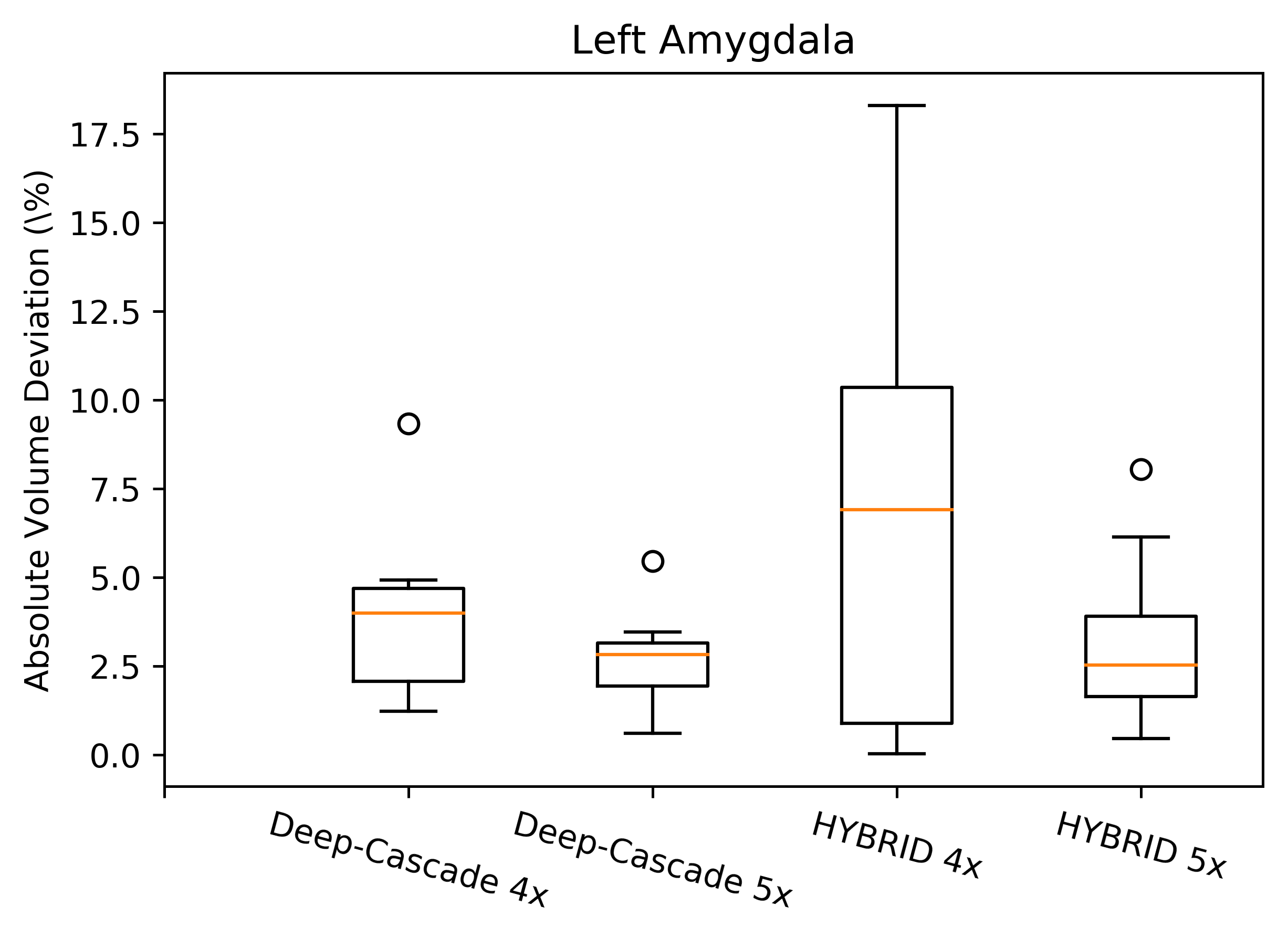}}
\subfloat[]{\includegraphics[width=0.25\textwidth]{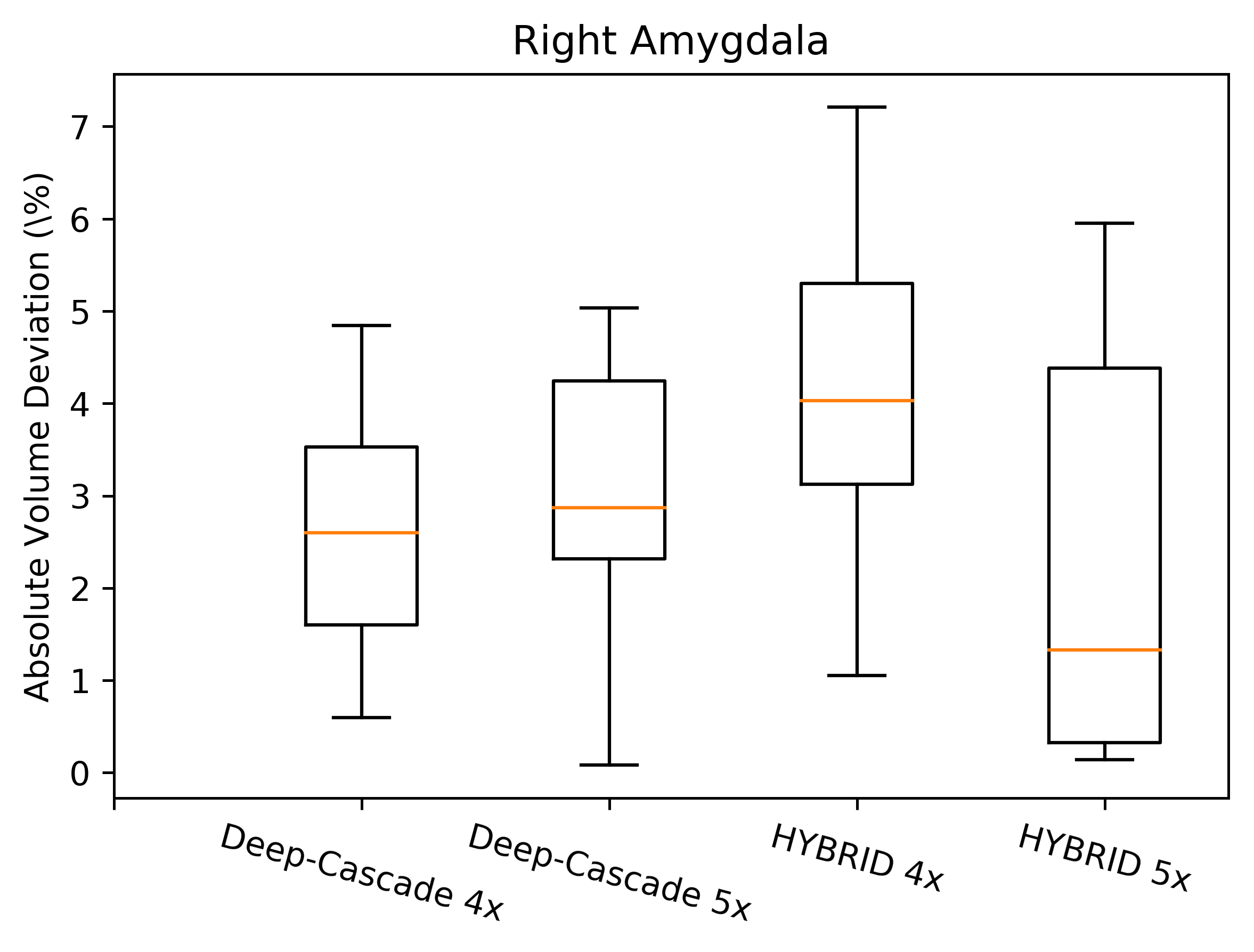}}\\
\caption{Summary of volume analysis processing results.}
\label{box_plots}
\end{figure*}

Volumetric analysis results are summarized in Table \ref{table_freesurfer}. An example of a subject where processing failed for the fully sampled reconstruction and the Deep-Cascade network (for both $4\times$ and $5\times$ acceleration factors) is presented in Figure \ref{freesurfer_failed}. Box-plots of the estimated volumes of total intra-cranial volume, white-matter, gray-matter, hippocampus, and amygdala volumes are shown in Figure \ref{box_plots}. These plots represent the distribution of eight volumes from only eight subjects in the test set (because volumetric analysis failed to process two fully sampled image sets). Due to  the reduced sample size ($n=8$) and because of reduced statistical power, we did not perform statistical tests on the volumetric data.

\begin{figure*}[!ht]
\centering
\includegraphics[width=0.75\textwidth]{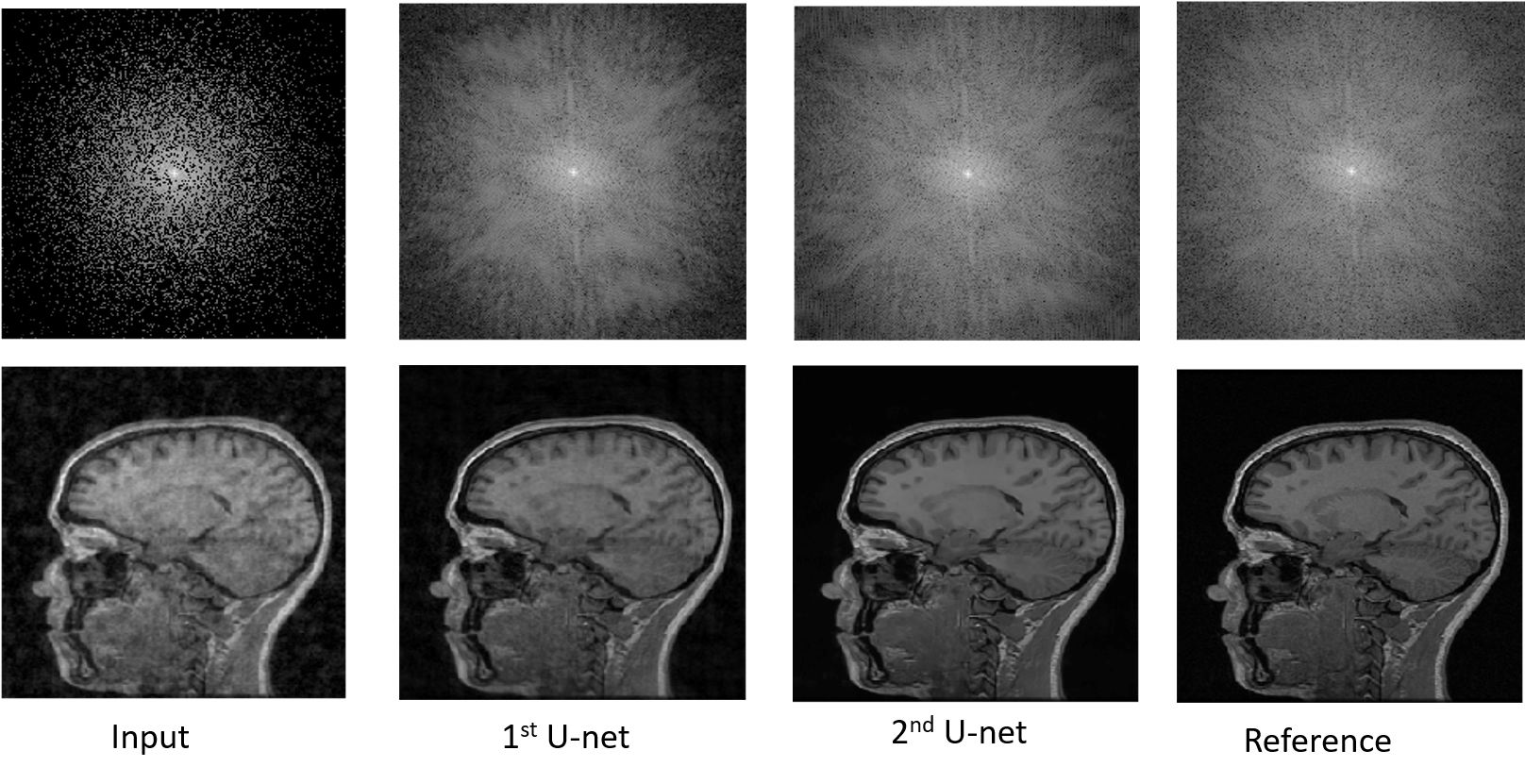}
\caption{Illustration of the images and their corresponding k-spaces at different stages of the Hybrid network. From the left to the right: Input undersampled k-space reconstruction ($NRMSE = 3.2\%$), result of the frequency domain U-net ($NRMSE = 1.9\%$), result of the image domain U-net ($NRMSE = 1.6\%$), and fully sampled reference reconstruction.}
\label{pipeline}
\end{figure*}

\begin{table*}[!t]
\caption{Summary of the results for the different architectures and different acceleration factors. The top two results for each metric and acceleration factor are emboldened.}
\label{table_results}
\centering
\begin{tabular}{c|c|c|c|c}
\hline
Acceleration factor & Model &  SSIM & NRMSE ($\%$) & PSNR \\ \hline
\multirow{4}{*}{$4\times$} 
& UNET &  $0.977 \pm 0.062$ & $2.326 \pm 1.039$ & $33.283 \pm 3.144$ \\
& DAGAN & $0.963 \pm 0.105$ & $2.925 \pm 1.474$ & $31.330 \pm 3.112$ \\
& RefineGAN & $0.979 \pm 0.068$ & $1.898 \pm 1.300$ & $35.436 \pm 3.705$ \\
& Deep-Cascade &  \boldmath{$0.986 \pm 0.054$} & \boldmath{$1.198 \pm 1.057$} & \boldmath{$39.510 \pm 3.345$}\\
& Hybrid   & \boldmath{$0.981 \pm 0.065$} & \boldmath{$1.664 \pm 1.041$} & \boldmath{$36.380 \pm 3.266$} \\ \hline
\multirow{4}{*}{$5\times$} 
& UNET   & $0.966 \pm 0.096$ & $2.727 \pm 1.174$ & $31.884 \pm 3.133$ \\ 
& DAGAN &  $0.949 \pm 0.110$ & $3.866 \pm 1.435$ & $28.691 \pm 2.658$ \\ 
& RefineGAN& $0.973 \pm 0.082$ & $2.273 \pm 1.401$ & $33.844 \pm 3.825$ \\ 
& Deep-Cascade & \boldmath{$0.982 \pm 0.068$} & \boldmath{$1.453 \pm 1.106$} & \boldmath{$37.668 \pm 3.202$}  \\ 
& Hybrid   & \boldmath{$0.978 \pm 0.080$} & \boldmath{$1.783 \pm 1.131$} & \boldmath{$35.772 \pm 3.214$}\\ \hline
\end{tabular}
\end{table*}


\begin{table}[!t]
\caption{Summary of volume analysis processing failures.}
\label{table_freesurfer}
\centering
\begin{tabular}{c|c|c}
\hline
 Model & Acceleration factor & \# failures (\%) \\ \hline
Fully sampled & $1\times$ & 2 (20\%) \\ \hline
\multirow{2}{*}{DEEP-CASCADE} 
& $4\times$ & 2 (20\%) \\ 
& $5\times$ & 1 (10\%) \\ \hline
\multirow{2}{*}{Hybrid} 
& $4\times$ & 0 (0\%) \\ 
& $5\times$ & 0 (0\%) \\ \hline
\end{tabular}
\end{table}

\section{Discussion}
Our hybrid method operates in frequency domain using a residual U-net and image domain using a U-net. The two networks are connected through the magnitude of the iDFT operation, and the model is fully trained end-to-end. An example for one subject of the input, intermediary and output results of our model are depicted in Figure \ref{pipeline}. Improvement in the image quality and NMRSE is noticeable at the output of the frequency domain network ($NRMSE = 1.9\%$), and further improvement can be seen at the output of the image domain network ($NRMSE = 1.6\%$). NMRSE for the input image was $3.2\%$.

Our Hybrid model achieved the second best metrics in the quantitative assessment, losing only to Deep-Cascade. RefineGAN was the third best method. It is important to mention that in the original RefineGAN paper the authors did not use a test set. They reported results on a validation set. The poorest performing technique was DAGAN, though it still had NMRSE $<4\%$ and SSIM $>0.94$. It is also important to highlight that in the original DAGAN paper, the authors did not use MR raw k-space data in their experiments. They computed the FFT of magnitude images followed by retrospective undersampling. This is not a realistic scenario, because when applying the FFT operator to magnitude images, the output is a k-space with Hermitian symmetry, while raw k-space is not Hermitian.

Visual assessment of the reconstructions (Figure \ref{reconstruction}) indicate that Hybrid and Deep-Cascade reconstruction are the best at preserving fine details as can be more noticeably seen in the cerebellum region. With the Hybrid reconstruction edges seem sharper, which might be an explanation to the fact that volumetric analysis software was able to successfully process all ten image dates sets, while it failed twice when processing the fully sampled reconstruction, and once with Deep-Cascade (acceleration factor of $4\times$). The volumetric analysis also failed to process another subject reconstructed with a Deep-Cascade reconstruction (acceleration factor of $5\times$). The quality of the brain structure segmentations vary according to the reconstruction. This is specially noticeable in the brain extraction step (Figure \ref{freesurfer_failed}). 

The deviation of the volumes estimated from the fully sampled reconstruction measurements were used as gold-standard (reference) measure for the eight successful subjects. Good agreement was found between the reference and volumetric results for Deep-Cascade and Hybrid image sets. For total intra-cranial volume the average absolute volume deviation was $<0.1\%$ in all subjects. For cerebral white-matter the absolute volume deviation was $<6\%$ in all subjects, and for gray-matter, it was $<4\%$ in all subjects. The average differences for the hippocampus and amygdala volumes was between $2\%$ and $3\%$. These results are a suggestive of the feasibility of incorporating a $5\times$ acceleration factor in a clinical or research imaging setting.

In terms of processing time, we did not perform a systematic comparison due to the fact that the different methods were implemented in different deep learning frameworks and many of the networks were not optimized. Nevertheless, all five techniques assessed take only a few seconds to reconstruct an entire MR volume using a cloud-based GPU (Amazon Elastic
Compute Cloud). In terms of processing times, UNET, DAGAN, Deep-Cascade and Hybrid took $<6$ hours to train, while RefineGAN training time was $\approx 72$ hours.  

\section{Conclusion}
In this work, we proposed a hybrid frequency domain/image domain CS MR reconstruction method that leverages the information of the iDFT mathematical formulation, essentially reducing the number of parameters of our model by orders of magnitude (for a $256\times256$ image, the Hybrid model required $10^3\times$ fewer coefficients compared to \cite{RN289}). Our model method was the second best in the quantitative comparison and it was the only one that did not fail in the volumetric analysis processing pipeline. Also, our Hybrid model produced the most visually pleasing images (Figure \ref{reconstruction}).

As future work, we would like to investigate the use of DC layers in our architecture and add an adversarial component for potentially improving reconstruction of high-frequency contents of the image. We will also extend our model to a more generic framework that can potentially deal with the parallel imaging scenario (\textit{cf.}, \cite{sense,grappa}).


%



\section*{Acknowledgment}
The authors would like to thank the Natural Science and Engineering Council of Canada (NSERC) for operating support, and Amazon Web Services for access to cloud-based GPU services. We would also like to thank Dr. Louis Lauzon for setting up the script to save the raw MR data at the Seaman Family MR Centre. R.S. was supported by an NSERC CREATE I3T Award and currently holds the T. Chen Fong Scholarship in Medical Imaging from the University of Calgary. R.F. holds the Hopewell Professor of Brain Imaging at the University of Calgary.

\ifCLASSOPTIONcaptionsoff
  \newpage
\fi



%
\bibliographystyle{IEEEtran}
\bibliography{sample}

%








\end{document}